# Design of dispersive optomechanical coupling and cooling in ultrahigh-$Q/V$ slot-type photonic crystal cavities

Ying Li[1,*], Jiangjun Zheng[1], Jie Gao[1], Jing Shu[1], Mehmet Sirin Aras[1], and Chee Wei Wong[1,*]

[1] *Optical Nanostructures Laboratory, Center for Integrated Science and Engineering, Solid-State Science and Engineering, and Mechanical Engineering, Columbia University, New York, NY 10027, USA*

[*]*Authors e-mail addresses*: yl2584@columbia.edu ; cww2104@columbia.edu

**Abstract:** We describe the strong optomechanical dynamical interactions in ultrahigh-$Q/V$ slot-type photonic crystal cavities. The dispersive coupling is based on a mode-gap photonic crystal cavities with light localization in an air mode with $0.02(\lambda/n)^3$ modal volumes while preserving optical cavity $Q$ up to $5 \times 10^6$. The mechanical mode is modeled to have fundamental resonance $\Omega_m/2\pi$ of 460 MHz and a quality factor $Q_m$ estimated at 12,000. For this slot-type optomechanical cavity, the dispersive coupling $g_{om}$ is numerically computed at up to 940 GHz/nm ($L_{om}$ of 202 nm) for the fundamental optomechanical mode. Dynamical parametric oscillations for both cooling and amplification, in the resolved and unresolved sideband limit, are examined numerically, along with the displacement spectral density and cooling rates for the various operating parameters.

## 1. Introduction

It is well-known that light has mechanical effects [1] and its radiation forces can be used to manipulate small atoms and particles [2]. Nowadays, the effects of optical forces in various mechanical and optical structures and systems have attracted intense and increasing interest for investigation [3]. Especially, the field of cavity optomechanics develops very fast [4-7], with recent studies covering a vast span of fundamental physics and derived applications [8-28]. In this field, the optomechanical coupling between the supported mechanical and optical cavity modes are of key importance due to its direct relevance to the generated optical forces, and one main goal of the developed techniques is to cool the targeted mechanical mode to its quantum mechanical ground state [10, 20, 24, 27]. Several classes of cavity optomechanical systems have been explored. One of the initial efforts examines macroscopic movable mirrors in the Laser Interferometer Gravitational Wave Observatory (LIGO) project [29-30]. Based on the micro- and nano-fabrication techniques, optomechanical resonators such as mirror coated AFM-cantilevers [14], movable micromirrors [15-16], vibrating microtoroids [11, 31], and nano-membranes [21,32] have been examined recently. Radiation-pressure dynamic backaction could be observed in these geometries. In addition, another class of optomechanical devices utilizes optical gradient forces [33-38] based on near-field effects. Compared to radiation-pressure based optomechanical cavities, these devices can achieve wavelength-scale effective optomechanical coupling lengths due to the strong transverse evanescent-field coupling between the adjacent cavity elements [25, 26, 33, 34, 35, 18, 38]. Photonic crystal membranes can be a very good candidate platform with great design flexibility [39-44], with photonic crystal cavities offering an ultrahigh optical quality factor with a small volume [45-47]. The internal optical intensity is very high and sensitive to the geometrical changes. However, to make these cavities support mechanical cavity modes with strong coupling with the optical modes, special design considerations are needed. Current reported geometries are either in-plane in side-by-side configuration [48-49] or vertically superimposed in face-to-face configuration [50]. Both configurations are recently examined experimentally to be promising for cavity optomechanical applications.

In this paper, we theoretically investigate the large dispersive optomechanical coupling between the mechanical and optical modes of a tuned air-slot mode-gap photonic crystal cavity [51]. First, the optical modes are shown to exhibit high optical quality factor ($Q$) with ultra-small modal volumes ($V$) [52-56], from three-dimensional finite-difference time-domain numerical simulations. The mechanical modes and properties are then modeled using finite element methods. Based on first-order perturbation theory [57-58] and parity considerations, the respective optomechanical modes are then examined numerically. The dynamical backaction of slot-type photonic crystal cavities are studied, including the optically-induced stiffening, optical cooling and amplification, and radio-frequency spectral densities, for various laser-cavity detuning, pump powers and other operating parameters. We also note that the slot-type photonic crystal cavity can operate in the resolved-sideband limit, which makes it possible to cool the mechanical motion to its quantum mechanical ground state.

**2. Optomechanical slot-type cavity design**

*2.1 Ultrahigh-Q/V cavity optical modes*

The slot-type optomechanical cavity is based on the air-slot mode-gap optical cavities recently demonstrated experimentally for gradual width-modulated mode-gap cavities [51] or heterostructure lattices [54], and theoretical proposed earlier in Ref. [47]. A non-terminated air-slot [55] is added to width-modulate line-defect photonic crystal cavities to create ultrasmall mode volume cavities. To better understand the various modes existing in the air-slot mode-gap cavities, the modes in the slotted photonic crystal waveguide with W1 line-defect width and their dispersion properties are first investigated and shown in Fig. 1(a) for the three localized waveguide modes. Mode I and II can be traced back to the W1 waveguide fundamental even mode and high-order odd mode respectively inside the photonic band gap, while mode III can be understood as arising from the second index-guided mode (as shown in Ref. [59]) below the projected bulk modes. We produce the cavities by locally shifting the air holes away from the center of waveguide – thus the cavity mode resonances are created below the transmission band of the slotted waveguide. Two of the possible modes in the cavities are shown in Fig. 1(b). Confirmed from the mode frequency and symmetry, cavity mode I is due to the mode gap of slotted waveguide mode I [Fig. 1(b)] and is expected to have both high $Q$ and sub-wavelength $V$. Cavity mode II [Fig. 1(c)] represents the mode with the same odd symmetry as mode II in slotted waveguide.

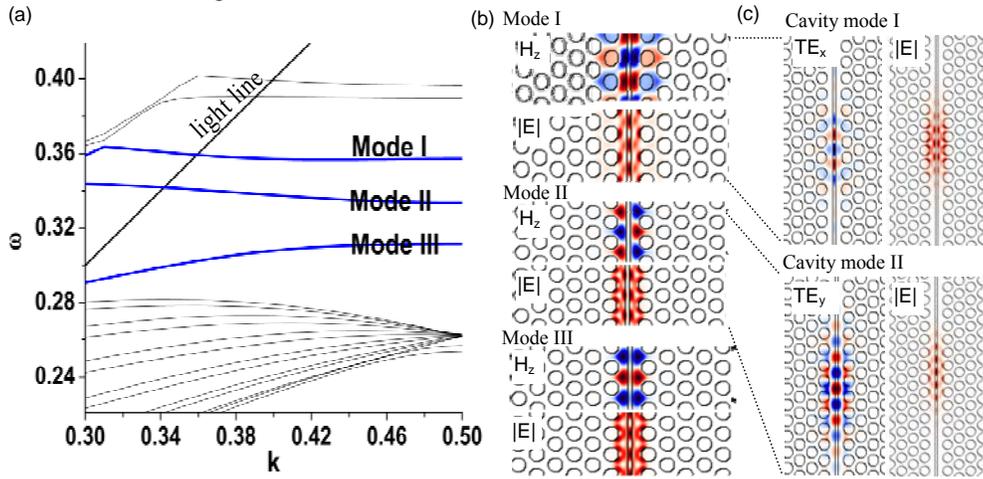

Fig. 1. (a) Photonic band structure of slotted PhCWG with $s$=80nm. The blue lines show the three modes in the slotted PhCWG. (b) $H$-field and energy distribution of waveguide modes I, II and III. (c) $E$-field and energy distribution of the first (above) and the second (below) cavity modes.

A scanning electron micrograph (SEM) image of the cavity is illustrated in Fig. 2(a) with $a$=490nm, $r$=0.34a, $t$=0.449a, $n_{si}$=3.48, $s$=80nm, $d_A$=0.0286$a$, $d_B$=0.019$a$ and $d_C$=0.0095$a$. FDTD simulation is performed to numerically evaluate the properties of the cavity mode. Fig. 2(b) shows the measured radiation spectrum of the cavity. For $s$=80 nm, the air-slot mode-gap confined PCS photonic crystal nanocavity supports a high $Q$ localized even mode [Fig. 1(c)] with $Q$ factor up to $5\times10^6$ and a mode volume $V$ of 0.02 $(\lambda/n_{air})^3$ from numerical simulations [47, 51]. 2D Fourier transform of the electric field shows few leaky components inside the light cone, supporting the high $Q$ character of this air-confined mode. From Fig. 1(c), the optical field is mainly distributed in cavity region, and the simulation results also show that the minimum number of lateral lattice rows next to the cavity to maintain the high $Q$ is ~ three lateral lattice rows. We therefore designed each beam into three lines with eight holes in each line, $l$=8$a$.

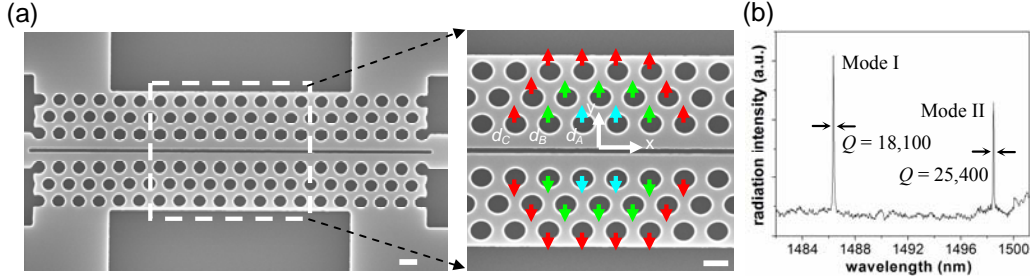

Fig. 2. (a) Example SEMs of the air-slot mode-gap optomechanical cavity. The holes shifts are shown in the right panel with $d_A$=0.0286$a$ (red), $d_B$=0.019$a$ (blue) and $d_C$=0.0095$a$ (green), where $a$ is the crystal lattice constant, for increasing the intrinsic cavity $Q$. Scale bar: 400nm. (b) Measured cavity radiation for the first two optical modes, with loaded $Q$ at 25,400 for the second mode and 18,100 for the first mode respectively.

*2.2 Cavity mechanical modes*

The mechanical modes are examined numerically via finite-element-method (FEM) simulations (COMSOL Multiphysics) for the dynamical motion of the suspended beams. The cavity mechanical modes can be categorized into common and differential modes of in-plane and out-of-plane motion [48] as well as compression and twisting modes of the two beams. The displacement fields $Q(r)$ of the first eight mechanical modes are shown in Fig. 3. In the numerical simulations, the beams are clamped at both ends using fixed boundary conditions at the two ends ($x=\pm1.96$um) of the beam, meanwhile limiting motion in the *x-y* plane (in boundary condition constraint of $z=\pm110$nm has a standard notion displacement in $R_z$=0nm and $R_x$=0nm, where $R_z$ ($R_x$) is the deformation along $z(x)$ axis), with silicon material properties: Young's modulus $E$ of 130GPa normal to [110] silicon crystallographic in-plane direction, thermal expansion coefficient $\alpha$ of $4.15\times10^{-6}$K, specific heat capacity $c$ of 703J/(kg·K), thermal conductivity $\kappa$ of 156W/(m·K) and density $\rho$ of 2330kg/m$^3$. We choose the triangular mesh configuration, with an average mesh element volume of ~ $9\times10^{-4}\mu m^3$, with the eigenfrequency and modal analysis for the first eight mechanical modes [Fig. 3], with eigenfrequencies ranging from 460 MHz to 2.16 GHz.

Only mechanical modes with parity $p_x = p_y = p_z = +1$ can coupled to the optical slot cavity modes due to the symmetry of the optical field, as described in Ref. [43]. Among the first eight mechanical modes, five of them (illustrated in grayscale in Fig. 4) do not couple to the optical modes due to parity considerations – *b* and *f* do not have the right parity in the *x* direction while *d*, *e* and *g* modes cannot be excited because of asymmetry of the optical gradient force along the *y* direction. In this slot cavity, therefore, only the first, third and eighth mechanical modes (depicted in color) have strong dispersive coupling to the localized optical modes. These are in-plane differential modes with modal frequencies $\Omega_m/2\pi$ at 459MHz, 1.36GHz and 2.16GHz respectively for a suspended beam length $L$ of 3.92um. The effective mass of each mode is computed from $m_{eff} = \int dV \frac{(r-r_0)^2}{(r-r_0)_m^2} \rho$, integrated over the computational space with $\rho$ defined as the mass density, $r$ the position from a fixed origin $r_0$ and $(r-r_0)_m$ defined as the maximum displacement. The effective mass of the first, third, and eighth mechanical modes are computed to be 200fg, 100fg, and 30fg respectively in our specific implementation with 3.92um beam length, width of 1.7um, and membrane thickness of 220nm.

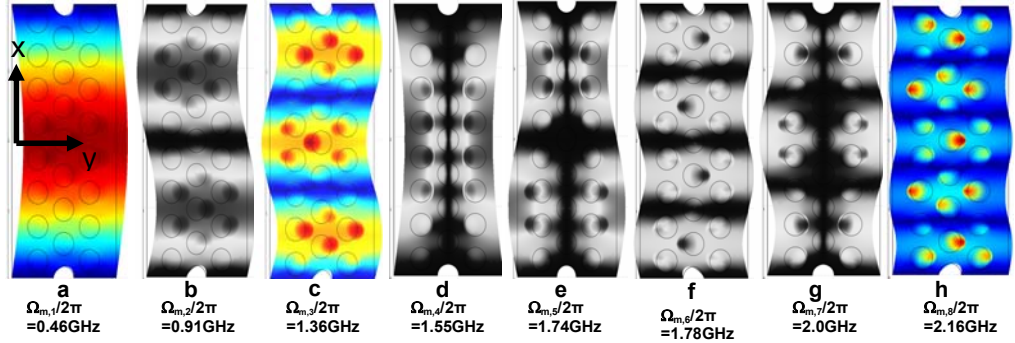

| a | b | c | d | e | f | g | h |
| --- | --- | --- | --- | --- | --- | --- | --- |
| $\Omega_{m,1}/2\pi$ =0.46GHz | $\Omega_{m,2}/2\pi$ =0.91GHz | $\Omega_{m,3}/2\pi$ =1.36GHz | $\Omega_{m,4}/2\pi$ =1.55GHz | $\Omega_{m,5}/2\pi$ =1.74GHz | $\Omega_{m,6}/2\pi$ =1.78GHz | $\Omega_{m,7}/2\pi$ =2.0GHz | $\Omega_{m,8}/2\pi$ =2.16GHz |

Fig. 3. Mechanical displacement profile of the first eight mechanical modes. Modes in color (*a*, *c*, *h*) are allowed by parity considerations to couple to the optical modes; modes in grayscale (*b*, *d*, *e*, *f*, *g*) are forbidden by parity for sizable optomechanical coupling. Red (blue) denotes maximum (minimum) displacement and plotted on a linear scale.

There are a number of possible dissipative processes where mechanical vibrational energy is dissipated into heat, either inside the structure or via interaction with its surroundings. These processes include squeezed film damping due to air viscosity [60], clamping losses, internal viscous damping in the silicon structure, and thermoelastic damping. Thermoelastic losses often set a lower ballpark estimate of the attainable $Q_m$ in a vibrating beam element, where $Q_{m,Zener}$ of the fundamental mechanical mode is expressed by [61,62]: $Q_{m,Zener} = \frac{c\rho}{E\alpha^2 T_R} \frac{1+(\omega\tau_z)^2}{\omega\tau_z}$, where $T_R$ is the ambient reservoir temperature, $\tau_z$ is the thermal relaxation time defined by $\frac{b^2}{\pi^2 \chi}$, $\chi = \frac{\kappa}{c\rho}$, and $b$ is the width of the beam. With silicon material properties, $T_R$ at 300K, and $b$ at 1.7 μm, $Q_m$ is found to be in the range of 12,000 for the fundamental mode, and 40,000 and 60,000 for the third and eighth mechanical modes respectively.

## 3. Coupling factor and symmetry considerations

Cavity optomechanics involves the mutual coupling of two modes in the same spatially co-located oscillator: one optical (characterized by its optical eigenfrequency and electromagnetic fields) and one mechanical (characterized by its mechanical eigenfrequency and displacement fields) degrees-of freedom. The perturbed cavity optical resonance, modified by small displacement about equilibrium displacement $\alpha$, can be given by its Taylor expansion around $\omega_o(\alpha)$. If we consider the first-order expansion, and also set $\omega_o(\alpha) = \omega_o|_{\alpha=\alpha_0}$ as the equilibrium resonance of the optical mode, then the first order $g_{om}=d\omega_o/d\alpha$ can be defined as optomechanical coupling rate. $g_{om}$ also represents the differential frequency shift of the cavity resonance ($\omega_o$) with mechanical displacement ($\alpha$) of the slot cavity beams. One can parameterize the interaction strength between optical and mechanical degrees-of-freedom by an effective coupling length $L_{om}$ [42] described by: $L_{om}^{-1} \equiv \frac{1}{\omega}\frac{d\omega}{d\alpha}$, with a corresponding optomechanical coupling frequency $g_{om}$ defined by $g_{om} \equiv \omega_o/L_{om}$.

*3.1 Perturbation theory*

Perturbation theory for Maxwell's equations with shifting material boundaries was used to calculate the coupling length $L_{om}$ [57, 58]. With the parameter $\Delta\alpha$ characterizing the perturbation, the Hellman-Feynman theorem [63] provides an exact expression for the derivative of $\omega$ in the limit of infinitesimal $\Delta\alpha$, $\frac{d\omega}{d\alpha} = -\frac{\omega^{(0)}}{2}\frac{\langle E^{(0)}|\frac{d\varepsilon}{d\alpha}|E^{(0)}\rangle}{\langle E^{(0)}|\varepsilon|E^{(0)}\rangle}$, where the terms with the (0) superscripts denote the unperturbed terms. With shifting material boundaries, the discontinuities in the $E$-field or the eigenoperator are overcome with anisotropic smoothening which gives the following expression for the integral in the numerator [57], $\langle E^{(0)}|\frac{d\varepsilon}{d\alpha}|E^{(0)}\rangle = \int dA \frac{dh}{d\alpha}\left[\Delta\varepsilon_{12}|E_\parallel^{(0)}|^2 - \Delta(\varepsilon_{12}^{-1})|D_\perp^{(0)}|^2\right]$, for first-order perturbation of the cavity resonance. The integral is performed across the entire boundary surfaces of the optomechanical cavity, with $h$ the displacement perpendicular to the unperturbed boundary surface, $\Delta\varepsilon_{12}$ defined as ($\varepsilon_1 - \varepsilon_2$) and $\Delta(\varepsilon_{12}^{-1})$ defined as ($\varepsilon_1^{-1} - \varepsilon_2^{-1}$). $|E_\parallel^{(0)}|^2$ is the unperturbed $E$-field parallel to the boundary surface while $|D_\perp^{(0)}|^2$ is the unperturbed electric displacement $D$ normal to the boundary surface. From Ref. [43], one defines $Q(r)=\alpha q(r)$, where $\alpha$ is the largest displacement amplitude that occurs anywhere for the displacement field $Q(r)$. From the perturbative formulation, one then obtains:

$$L_{OM}^{-1} = \frac{1}{2}\frac{\int dA\left(q(r)\cdot\hat{n}\right)\left[\Delta\varepsilon_{12}(r)|E_\parallel^{(0)}(r)|^2 - \Delta\left(\varepsilon_{12}^{-1}(r)\right)|D_\perp^{(0)}(r)|^2\right]}{\int dV\varepsilon(r)|E(r)|^2}$$, where $\hat{n}$ is the unit normal vector at the surface of the unperturbed cavity and the spatial $r$-dependence explicitly shown here.

*3.2 Optomechanical coupling in slot-type optical cavities*

Fig. 4 shows the computed optomechanical coupling in the slot-type mode gap cavities, from first-order perturbation theory. As noted from parity consideration, here we show the optomechanical coupling strengths for the first optical mode to the *allowed* first ($\Omega_{m,1}$) and second ($\Omega_{m,3}$) mechanical modes, denoted as $g_{om}(O_1\text{-}M_1)$ and $g_{om}(O_1\text{-}M_2)$ respectively. We illustrate the coupling strengths for different slot gaps $s$ of cavity, ranging from 40nm to 200nm. The electromagnetic field used is within a slot length $l=8a$, since the cavity is confined by the PhCWG mode gap to a spatial localization of only several lattice constants $a$. As shown in Fig. 4, when the first optical mode is coupled with the second mechanical mode, the $g_{om}$ is lower than that with the fundamental mechanical mode, which means the fundamental optical and mechanical modes provide the strongest dispersive coupling. The negative values depict as a decrease in optical resonance frequency for increasing slot widths $s$. For the fundamental mode, the dispersive coupling can go up to 940GHz/nm (or a coupling length of 202nm) for a slot width of 40nm. These strong optomechanical coupling is more than an order of magnitude larger than in earlier optomechanical implementations. We also note that, since the electromagnetic field is negligible outside the cavity region of $l = 8a$, the coupling length does not change much when $l$ is longer than $8a$.

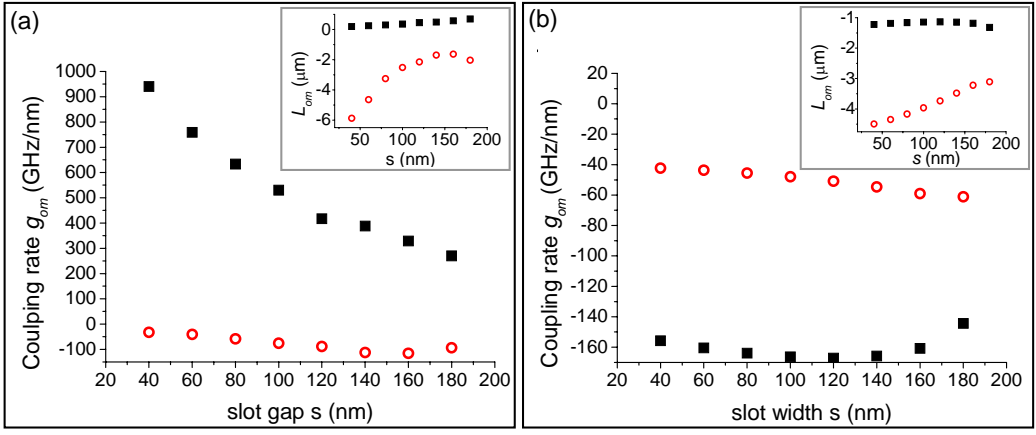

Fig. 4. (a) Computed optomechanical coupling rates of the fundamental optical mode with first (black solid squares) and second (red open circles) *allowed* mechanical modes, computed for the different slot gaps $s$. The inset panel is the corresponding coupling length. (b) Computed optomechanical coupling rates of the second optical mode coupled with first (black solid squares) and second (red open circles) *allowed* mechanical modes.

**4. Coupled mode theory**

The coupled equations of motion for the optical and mechanical modes can be derived from a single Hamiltonian [64, 8]: $\frac{da}{dt} = i\Delta(x)a - \left(\frac{1}{2\tau_0} + \frac{1}{2\tau_{ex}}\right)a + i\sqrt{\frac{1}{\tau_{ex}}}s$ and $\frac{d^2x}{dt^2} + \frac{\Omega_m}{2Q_m}\frac{dx}{dt} + \Omega_m^2 x = \frac{F_{OM}(t)}{m_{eff}} + \frac{F_L(t)}{m_{eff}} = -\frac{g_{OM}}{\omega_0}\frac{|a|^2}{m_{eff}} + \frac{F_L(t)}{m_{eff}}$,

where $|a|^2$ is the stored cavity energy, $|s|^2$ the launched input power into the cavity, with a cavity decay rate $\kappa$ of $\frac{1}{2\tau} = \frac{1}{2\tau_0} + \frac{1}{2\tau_{ex}}$, with intrinsic rate $1/\tau_o$ and coupling rate $1/\tau_{ex}$. $\Delta(x) = \omega - \omega_0(x)$ is the pump laser frequency $\omega$ detuning with respect to the cavity resonance $\omega_o(x)$ with explicitly displacement $x$ shown. In this case, we have $\Delta(x) = \Delta - g_{OM}x$. $F_L(t)$ is the thermal Langevin force. We illustrate the time-domain displacement $x(t)$ and the normalized cavity amplitude of the first optical and first mechanical modes in Fig. 5(a). The cavity amplitude oscillates in-phase with the displacement within the mechanical frequency cycle as shown. In Fig. 5(b) we show the optical cavity amplitude transduction for different normalized detunings ($\Delta\tau = -1, -0.25, 0, 0.25, 1$). At zero detuning and with a launched power $|s|^2$ into the cavity, the cavity amplitude oscillates with a single-period cycle at the fundamental mechanical mode frequency, as indicative of mixing of the optomechanical domains. At detunings $\Delta\tau = \pm 0.25$, a two-period cycle with a second amplitude maxima is distinctly observed, with inverted transmission between the blue and red detunings. At larger detunings (such as $\Delta\tau = \pm 1$), a two-period cycle is still observed, although the second amplitude maxima is suppressed.

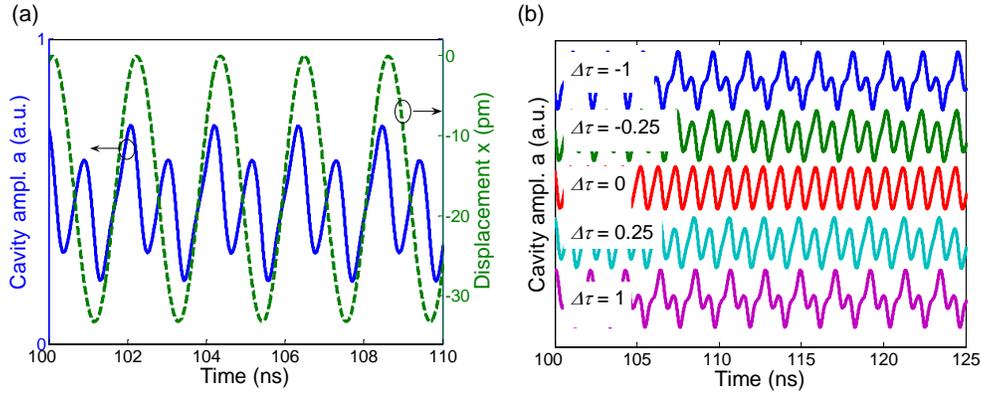

Fig. 5. (a) Time-domain cavity amplitude *a* (solid blue line) and displacement *x* (dashed green line) of the first optical and first mechanical modes, with $g_{om}$ of 940 GHz/nm, $\Omega_m/2\pi$ of 470 MHz, $Q_m$ of 12,400, $\kappa/2\pi$ of 425 GHz, and $(1/\tau_{ex})/2\pi$ of 38 MHz. (b) Time-domain cavity amplitude for normalized detunings $\Delta\tau$ at -1, -0.25, 0, 0.25 and 1 (top to bottom).

**5. Displacement spectral density**

*5.1 Optically-induced stiffening and effective damping rate*

From the coupled equations, the *x*-dependent contribution to this adiabatic response provides an optical contribution to the stiffness of the spring-mass system. The corresponding change in spring constant leads to a frequency shift relative to the unperturbed mechanical oscillator eigenfrequency, or termed as optically-induced stiffening [4, 48]. The non-adiabatic contribution in coupled equations is proportional to the velocity of the spring-mass system. The optical gradient force induced damping rate modifies the intrinsic mechanical resonator loss rate $\Gamma_m$, yielding an effective damping rate: $\Gamma_{eff} = \Gamma + \Gamma_m$, where

$$\Gamma = -\frac{\omega_0}{2\Omega_m L_{om}^2 m_{eff}}\left(\frac{2\kappa_{ex}}{\kappa^2 + 4\Delta^2}\right)\left[\frac{\kappa/2}{(\Delta-\Omega_m)^2+(\kappa/2)^2} - \frac{\kappa/2}{(\Delta+\Omega_m)^2+(\kappa/2)^2}\right]P.$$

We note that this is valid only in the weak retardation regime in which $\kappa \gg \Omega_m$. We illustrate in Fig. 5 the corresponding frequency shifts and effective damping rate of the slot-type mode-gap cavity, for different input powers and normalized detuning ($\Delta\tau$). With this classical model, the laser introduces a damping without introducing a modified Langevin force. This is a key feature and allows the enhanced damping to reduce the mechanical oscillator temperature, yielding as a final effective temperature $T_{eff}$ for the mechanical mode under consideration: $T_{eff} \cong \frac{\Gamma_m}{\Gamma_{eff}} T_R$. As shown in Fig. 5, as optical *Q* increases, at certain detuning the frequency shift becomes larger and the effective temperature is lowered, denoting the increased cooling rate. For a fixed optical *Q* in the unresolved sideband limit, there will be an optimal detuning where the linewidth reaches its largest value and the effective temperature is the lowest. In our case this optimal detuning $\Delta\tau$ is around -0.25 with an input power of 50pW and the effective temperature can be lower than 50K.

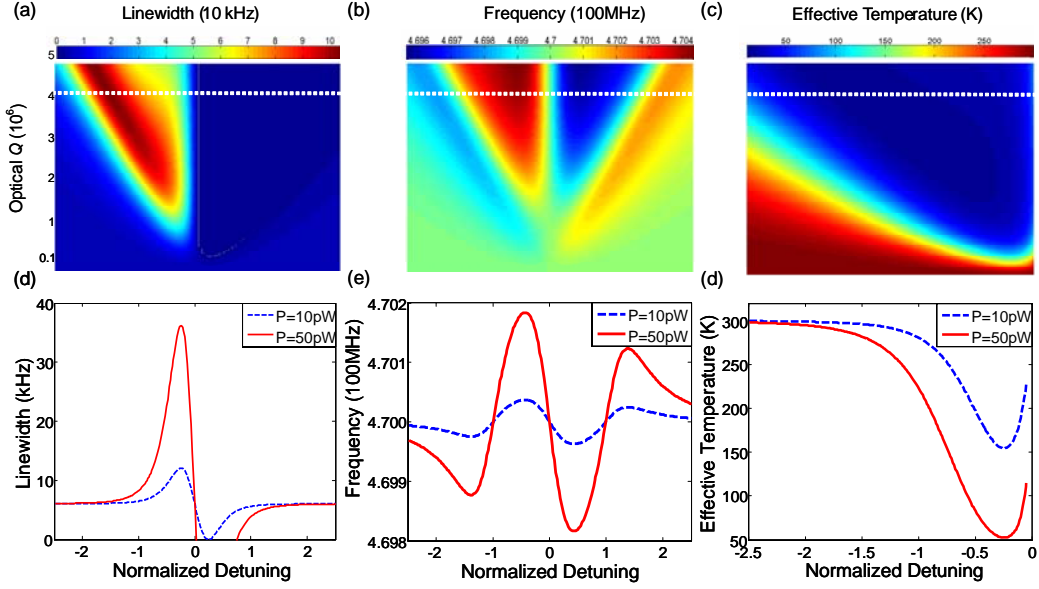

Fig.6. (a-c) Two-dimensional surface plots of the first optical – first mechanical mode linewidth (a) mechanical frequency (b) and effective temperature (c), for varying detunings and optical $Q$ factors. A fixed pump power of 1pW is used, along with an effective mass of 200fg and a 300K bath temperature. The dashed white line denotes the condition for $\Omega_m = \kappa$. (d-f) Example first optical – first mechanical mode linewidths (d), frequency shift (e) and effective temperature (f) with two input powers ($P$) and varying laser-cavity detuning. Otherwise indicated, the conditions are identical to panel (a), and with optical $Q$ chosen at $5\times10^5$.

The spectral intensity of purely mechanical displacement in the oscillator is described as: $S_x(\Omega) = \dfrac{2\Gamma_m k_B T / m_x}{\left(\Omega_m^2 - \Omega^2\right)^2 + \left(\Omega\Gamma_m\right)^2}$, without the optical stiffening and damping. Since the coupling will shift the oscillator frequency and damping, we can modify $\Omega_m$ and $\Gamma_m$ in the expression into $\Omega_m^{'} = \Omega_m + \Delta\Omega_m$ and $\Gamma_m^{'} = \Gamma + \Gamma_m$. Fig. 7(a) shows the resulting displacement spectral density when the input power $P$ changes from 0 to 6.9uW, and normalized detuning $\Delta\tau$=-0.25 where the linewidth has the maximum value and the frequency shift is positive. With increasing input power, the peak value of the displacement spectral density goes down and the full-width at half-maximum becomes larger, which demonstrates an effective cooled temperature of the slot-type optomechanical oscillator. In Fig. 7(b) we show the optical stiffing and linewidth damping of the first two mechanical modes, for a span of detunings while maintaining a fixed input power. Note that the optical stiffening is not monotonic with increasing detuning. For a cavity decay $\kappa/2\pi$ of 387 MHz, the optimal detuning is at $\Delta\tau$ of -0.43, for the largest optical gradient force stiffening. For the second allowed mode, in the region of normalized detuning from zero to -4, this stiffening is large which leads to a significantly suppressed spectral density. Moreover, note that in both Fig. 7(a) and 7(b), a large optical stiffening can be observed in the slot-type optomechanical cavity, where the optical stiffening can result in a modified mechanical frequency more than 1.86× the bare mechanical frequency.

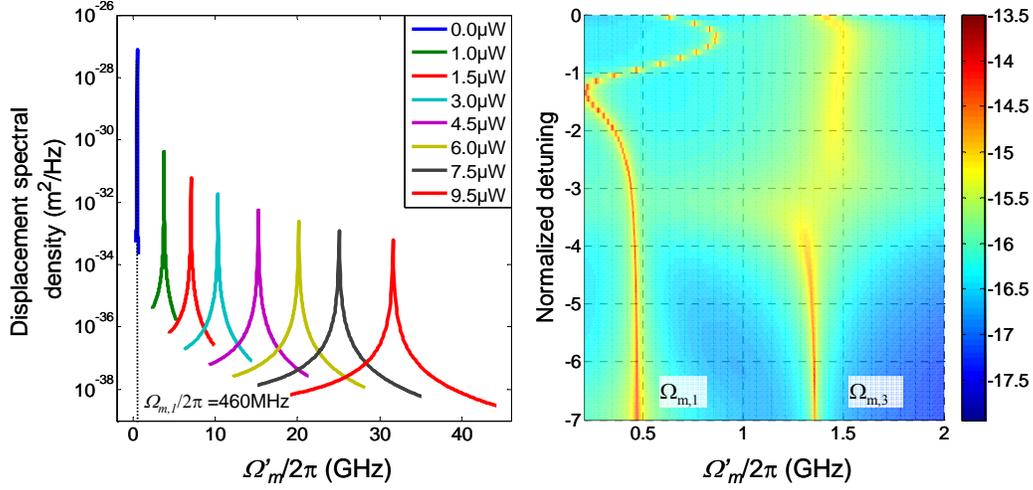

Fig. 7. (a) Displacement spectral density of the first mechanical mode, with optical detuning from the first optical mode. With the input power increasing from 0 to 9.5uW, in addition to an observed optical stiffening, the amplitude decreases with a larger linewidth for a decrease in the effective temperature. The detuning $\Delta\tau$ is fixed at -0.25, for an optical $Q$ of $5\times10^5$, $m_{eff}$ of 200fg, at 300K bath temperature. (b) Displacement spectral density of the first and second allowed mechanical modes with different detunings. The scale bar is in dB with units of m$^2$/Hz (pump powers $P_1$ of 0.1uW and $P_2$ of 50uW used respectively in the modeling).

As shown above, both cooling and amplification can be realized in the optomechanical cavity through the red- and blue-detuning to the cavity resonance. An important question is what limiting temperature is achievable with the optical gradient force backaction cooling technique as described above. Two theoretical papers [22, 26] have extended the classical theory of radiation-pressure backaction cooling to the quantum regime and shown the close relationship that cavity backaction cooling has with the laser cooling of harmonically bound atoms and ions. The result can be simply divided by two conditions. In the unresolved side-band regime, $\kappa \gg \Omega_m$, the ground state cooling is limited as: $n_f \approx \frac{\kappa}{4\Omega_m} \gg 1$, where $n_f$ is the minimum phonon number. On the other hand, in the resolved side-band regime, $\Omega_m \gg \kappa$, occupancies well below unity can be attained yielding: $n_f \approx \frac{\kappa^2}{16\Omega_m^2} \ll 1$. Most of the present optomechanical cavities are in the unresolved sideband regime, either because low optical quality factor or low mechanical frequency, which limit the minimum phonon number higher than unity. However, since our ultrahigh-$Q/V$ slot-type photonic crystal cavity has a high optical $Q$ factor and higher mechanical frequency due to its small volume, it has significant potential to operate into the resolved sideband region. For example, for the first mechanical mode ($\Omega_m/2\pi$ of 460 MHz), an optical $Q$ of more than $5\times10^5$ will bring the optomechanical oscillator within the resolved sideband limit with a $n_f$ of $1\times10^{-3}$, allowing the potential to cool the mechanical mode to its ground state.

### 6. Conclusion

We illustrate numerically the slot-type mode-gap photonic crystal cavities for strong optical gradient force interactions. With the simultaneous strong optical field localization in $0.02(\lambda/n)^3$ modal volumes and cavity $Q$s up to $5\times10^6$, we examined the optomechanical transduction of the various mechanical and optical modes for a dispersive coupling $g_{om}$ up to 940 GHz/nm for the fundamental modes. Temporal coupled oscillations between the optical and mechanical fields are examined, along with effects of large optically-induced stiffening, cooling and resulting displacement spectral densities, for the various operating regimes in the slot-type optomechanical cavities.

**Acknowledgments**

The authors acknowledge discussions with S. Nellaiappan and P. Hsieh. This work is supported by DARPA DSO with program manager Dr. J. R. Abo-Shaeer under contract number C11L10831.